\documentclass[journal=jacsat,manuscript=article]{achemso}

\usepackage{chemformula} % Formula subscripts using \ch{}
\usepackage[T1]{fontenc} % Use modern font encodings

\author{Jaime Dolado}
\affiliation{European Synchrotron Radiation Facility, 38043 Grenoble, France}
\alsoaffiliation{Departamento de Física de Materiales, Facultad de Ciencias Físicas, Universidad Complutense de Madrid, 28040-Madrid, Spain}
\email{jaime-jose.dolado-fernandez@esrf.fr}

\author{Fernanda Malato}
\affiliation{Instituto de Ciencia de Materiales de Madrid, Consejo Superior de Investigaciones Científicas, 28049 Cantoblanco, Spain}

\author{Jaime Segura-Ruiz}
\affiliation{European Synchrotron Radiation Facility, 38043 Grenoble, France}

\author{María Taeño}
\affiliation{Departamento de Física de Materiales, Facultad de Ciencias Físicas, Universidad Complutense de Madrid, 28040-Madrid, Spain}

\author{Irina Snigireva}
\affiliation{European Synchrotron Radiation Facility, 38043 Grenoble, France}

\author{Pedro Hidalgo}
\affiliation{Departamento de Física de Materiales, Facultad de Ciencias Físicas, Universidad Complutense de Madrid, 28040-Madrid, Spain}

\author{Bianchi Méndez}
\affiliation{Departamento de Física de Materiales, Facultad de Ciencias Físicas, Universidad Complutense de Madrid, 28040-Madrid, Spain}

\author{Gema Martínez-Criado}
\affiliation{European Synchrotron Radiation Facility, 38043 Grenoble, France}
\alsoaffiliation{Instituto de Ciencia de Materiales de Madrid, Consejo Superior de Investigaciones Científicas, 28049 Cantoblanco, Spain}

\title[An \textsf{achemso} demo]
  {Zn$_2$GeO$_4$/SnO$_2$ nanowire architecture: A correlative spatially resolved X-ray study}


\begin{document}


\begin{tocentry}

\includegraphics[width=8cm]{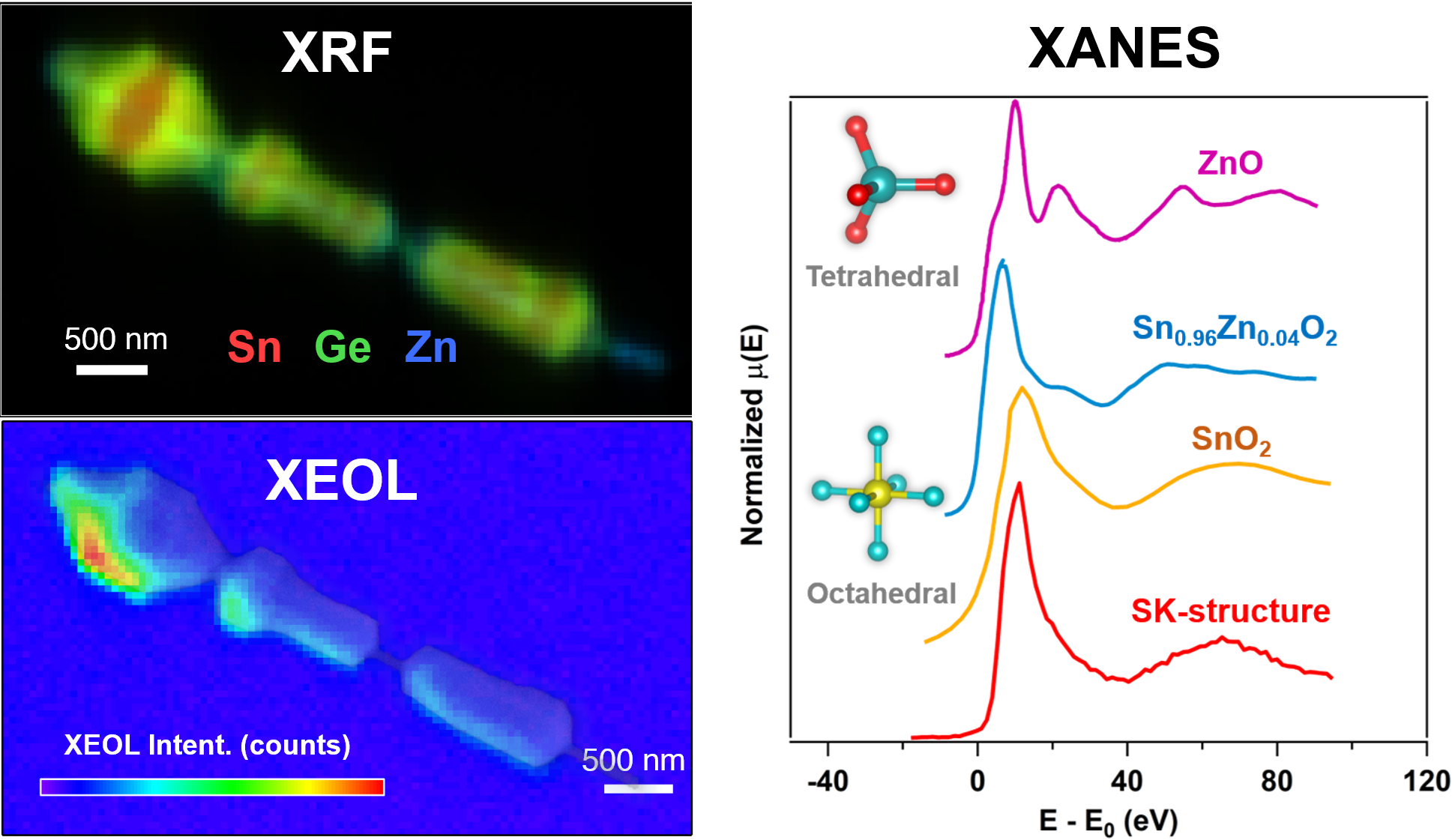}
\vspace{2cm}

\textbf{Zn$_2$GeO$_4$/SnO$_2$ nanowire architecture: A correlative spatially resolved X-ray study}

\vspace{0.5cm}
Jaime Dolado, Fernanda Malato, Jaime Segura-Ruiz, María Taeño, Irina Snigireva, Pedro Hidalgo, Bianchi Méndez, and Gema Martínez-Criado

\vspace{1cm}
\textbf{Synopsis}\\
Herein we present results on a correlative hard X-ray study of Zn$_2$GeO$_4$/SnO$_2$ nanowire heterostructures formed by Plateau-Rayleigh instability. The multi-technique synchrotron approach allows unraveling the elemental composition, optical processes and short-range order, making it possible to determine them independently and spatially resolved at the nanometer length scale. This study highlight the role of diffused impurities on the optical properties of nano-heterostructures and open new avenues for further heterogeneity studies using the X-ray nanoprobe.

\end{tocentry}

\begin{abstract}
Wide bandgap semiconductor heterostructures based on nanowires have attracted great interest as one of the nanoscale building blocks for the development of novel optoelectronic devices. Owing to the critical role of heterogeneities and structural imperfections in the optoelectronic properties, it has become a challenge to investigate in detail the local structure and elemental composition of these systems. Typically, transmission electron microscopy and energy dispersive X-ray spectroscopy are applied to address such issues. Here, using a synchrotron nanobeam probe, we show a correlative spatially-resolved study of the composition, luminescence and local structure of a Zn$_2$GeO$_4$/SnO$_2$ nanowire-based system. Unlike other works, the heterostructures were synthesized by an effective catalyst-free thermal evaporation method and present a skewer-shaped architecture formed by Plateau-Rayleigh instability. Our results highlight the role of diffused impurities on the optical properties of the heterostructure and open new avenues for further heterogeneity studies using the X-ray nanoprobe.
\end{abstract}


Nanowires represent key building blocks for high density device-based nanotechnology. They are critical elements to implement the “bottom-up” approach to nanoscale fabrication, whereby well-defined nanosized architectures with unique physical properties are assembled into electronic and photonic devices. Moreover, by manipulating growth kinetics it is possible today to create more complex nanoscale heterostructures. The introduction of sequential branching, for example, with variations in the composition and/or doping will make possible novel electronic and photonic functionality into these building blocks\cite{lauhon2002, lieber2003, jia2019}. Likewise, owing to the Plateau-Rayleigh (PR) instability, for example, Zn$_2$GeO$_4$/SnO$_2$ heterostructures with skewer-like (SK) morphologies have been recently synthesized\cite{dolado2019}, emerging as direct competitors to traditional wide bandgap materials for ultraviolet optoelectronic nanodevices. Thus, elongated Zn$_2$GeO$_4$ nanorods (E$_g\sim$ 4.5 eV) decorated with a linear array of SnO$_2$ crystals (E$_g\sim$ 3.6 eV) in a necklace fashion were made by an effective thermal evaporation method. Yet, the concept underlying these building blocks rests on controlling the structural and optical properties at the nanometer scale. Often, the oxide nano-heterostructures suffer from native defects, heterogeneities, and/or local structural imperfections that can drastically affect their optoelectronic response\cite{dayeh2011, huijben2013, li2016}. So far, the structural and compositional quality of these nanowire architectures has been mostly gauged by correlative transmission electron microscopy (TEM) and energy dispersive X-ray spectroscopy (EDS) measurements\cite{dolado2019, jung2007, liu2008, lim2009, gao2019}. Additionally, indirect methods such as cathodoluminescence measurements have been also applied to determine local heterogeneities in these nanowire-based systems\cite{hetzl2016, songmuang2016}. However, the formation of the individual components of these decorated nanostructures also needs a systematic structural and chemical study at the nanometer scale. Although the creation of crystalline phases driven by PR mechanism has been demonstrated, so far there are many open questions. Just few examples include the role of impurities on coupling formation and local atomic site configuration\cite{xu2018}, how these necklace structures are joined to form decorated architectures\cite{day2016, li2017, chu2018}, whether there is a full control over composition, diffusion paths, and/or structural modifications\cite{day2015, xue2016, liu2020}, coexisting structural polytypes\cite{gorshkov2020, walbert2020}, diffusion-driven effects\cite{garin2017}, phase separation problems\cite{allaire2021}, radial or axial chemical modulations\cite{walbert2020}, as well as atomic scale variability due to fluctuations in the growth rate\cite{dolado2019}.

Consequently, apart from a few cases in which single crystal X-ray diffraction measurements of nanowires could be recorded\cite{bussone2015, li2020, mostafavi2021}, almost no information about the crystal symmetry of individual nanostructures and their integration into single architectures is available. Some works have reported the first applications of nano-X-ray absorption spectroscopy to single nanowires\cite{segura2011, segura2014, martinez2014}. Thus, if X-ray excited optical luminescence is added to an X-ray absorption experiment, the challenge of a correlative study of single nanostructures can be overcome using hard X-ray nanobeams. Therefore, here we address such issues by characterizing a representative Zn$_2$GeO$_4$/SnO$_2$ nano-heterostructure formed by PR mechanism obtained in a one-step thermal evaporation method\cite{dolado2019}.

In this work we illustrate the use of a hard X-ray nanoprobe to study decorated nanowire based architectures by multiple contrast mechanisms using the nano-analysis beamline ID16B of the European Synchrotron Radiation Facility (ESRF). Using a pair of Kirkpatrick-Baez Si mirrors, our study involves the collection of X-ray fluorescence (XRF) and X-ray excited optical luminescence (XEOL) emissions induced by a highly focused and intense hard X-ray nanobeam. Both XRF and XEOL signals were recorded simultaneously. Whereas the characteristics XRF is recorded with an energy dispersive Si drift detector (deadtimes below 20\%) at an angle of $\theta$ = (15 ± 5)º with respect to the sample surface, the luminescence is detected normal to this surface by a far-field optical collection system\cite{martinez2016}. We apply this experimental scheme to SK-structures composed by a Zn$_2$GeO$_4$ nanowires and tin oxide crystallites produced by an effective thermal evaporation method \cite{dolado2019}.  Compared to most of the previous reports\cite{jia2019}, here we use a catalyst-free mechanism to form the wide bandgap heterostructures. The droplets created by the PR instability act as nucleation centres, giving rise to an unprecedented control over the surface morphology\cite{day2015, day2016}.

\begin{figure}
    \centering
        \includegraphics[width=1\textwidth]{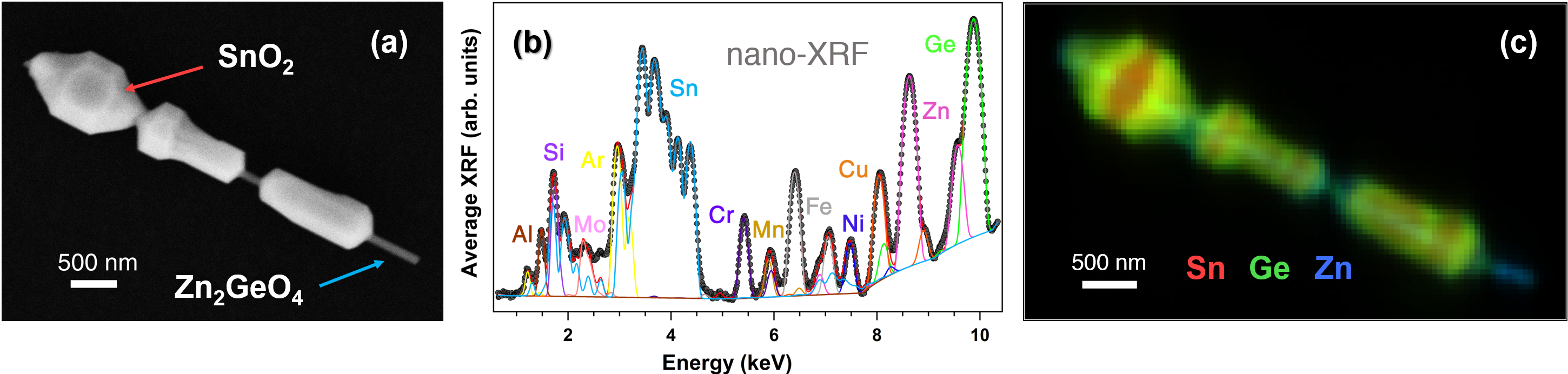}
        \caption{(a) SEM image of SK-structure. (b) Average XRF spectrum recorded over the SK- structure shown in (a) at 11.25 keV. (c) XRF map in RGB visualization that depicts the XRF intensities of Sn (red), Ge (green), and Zn (blue). Their colour brightness (light represents high counts, dark low counts) indicates the intensity ranges (more details in the Supporting Information).} 
        \label{fig1}
\end{figure}

Figure~\ref{fig1}a shows the scanning electron microscopy (SEM) image of a single Zn$_2$GeO$_4$/SnO$_2$ nanoheterostructure taken with a LEO 1530 SEM equipment. It consists of a straight central wire surrounded by well-faceted crystallites (SK-like shape). The wire dimensions are 5 $\mu$m length and 100 nm diameter, whereas the crystallites present sizes ranging between 500 - 800 nm. Figures~\ref{fig1}b-c shows the X-ray nanoimaging results obtained by raster-scanning the SK-structure in the X-ray nanobeam. The monochromatic X-ray beam was focused to 80 x 60 nm$^2$ spot size ($V$ × $H$) with $\sim$1011 ph/s at room temperature in air. The XRF maps were taken at 11.25 keV (above the absorption edge of Ge located at 11.1031 keV) with a pixel size of 50 x 50 nm$^2$ over a 3 × 5 $\mu$m$^2$ sample area with an integration time of 1000 ms per point. In addition to the major elements (Zn, Ge and Sn) of the Zn$_2$GeO$_4$/SnO$_2$ heterostructure, the average XRF spectrum (Figure~\ref{fig1}b) also reveals unintentional dopants (e.g., Cr, Fe, Ni and Cu) homogeneously distributed and likely originated from contaminants in precursors or in the reactor. Figure~\ref{fig1}c shows Zn (blue), Ge (green) and Sn (red) superimposed distributions (individual elemental distribution maps can be seen in the Supporting Information). The composition of the nanowire axis is basically Zn$_2$GeO$_4$, as expected. However, the XRF maps of the SnO$_2$ crystallites exhibit an interesting sandwich like pattern. A significant incorporation of Ge takes place within the SnO$_2$ crystallites, which could modify the short-range order and give rise to secondary phases or defects, affecting accordingly the carrier recombination mechanisms. Moreover, an axial variation of the Sn localization is observed within the crystallites, being especially high in the fully faceted areas, whereas Ge is particularly distributed in the regions not completely faceted (see Supporting Information for further details). Finally, the Zn signal (Figure SI-2c in Supporting Information) comes mainly from the nanowire with a much lesser degree incorporated into the crystallites.

In good agreement with these results, we recently reported the observation of a high Ge fraction within the SnO$_2$ crystallites (from 8\% to 17\%) by TEM-EDS. Our results point to the formation of a solid solution, such as Sn$_{1-x}$Ge$_x$O$_2$, which keeps the rutile crystal structure\cite{dolado2019}. Likely, the faceted areas present a lower amount of Ge (and a higher amount of Sn) because the Ge out-diffusion to the Zn$_2$GeO$_4$ nanowire surface (which is arranged in periodic droplets due to PR instability) acts as nucleation sites for the Sn crystals. The more faceted the crystal is, the more Sn replaced Ge to form the rutile crystal phase. Recent studies have shown that the nanowire coating in the PR-type processes is determined by parameters such as the evolution of the internal crystalline nanostructure over time, or the exchange of free atoms between the nanowire surface and the enveloping near-surface layer\cite{walbert2020, gorshkov2020}. Therefore, these factors could be involved in the different distribution of Ge and Sn in the SK-structure.

\begin{figure}
    \centering
        \includegraphics[width=1\textwidth]{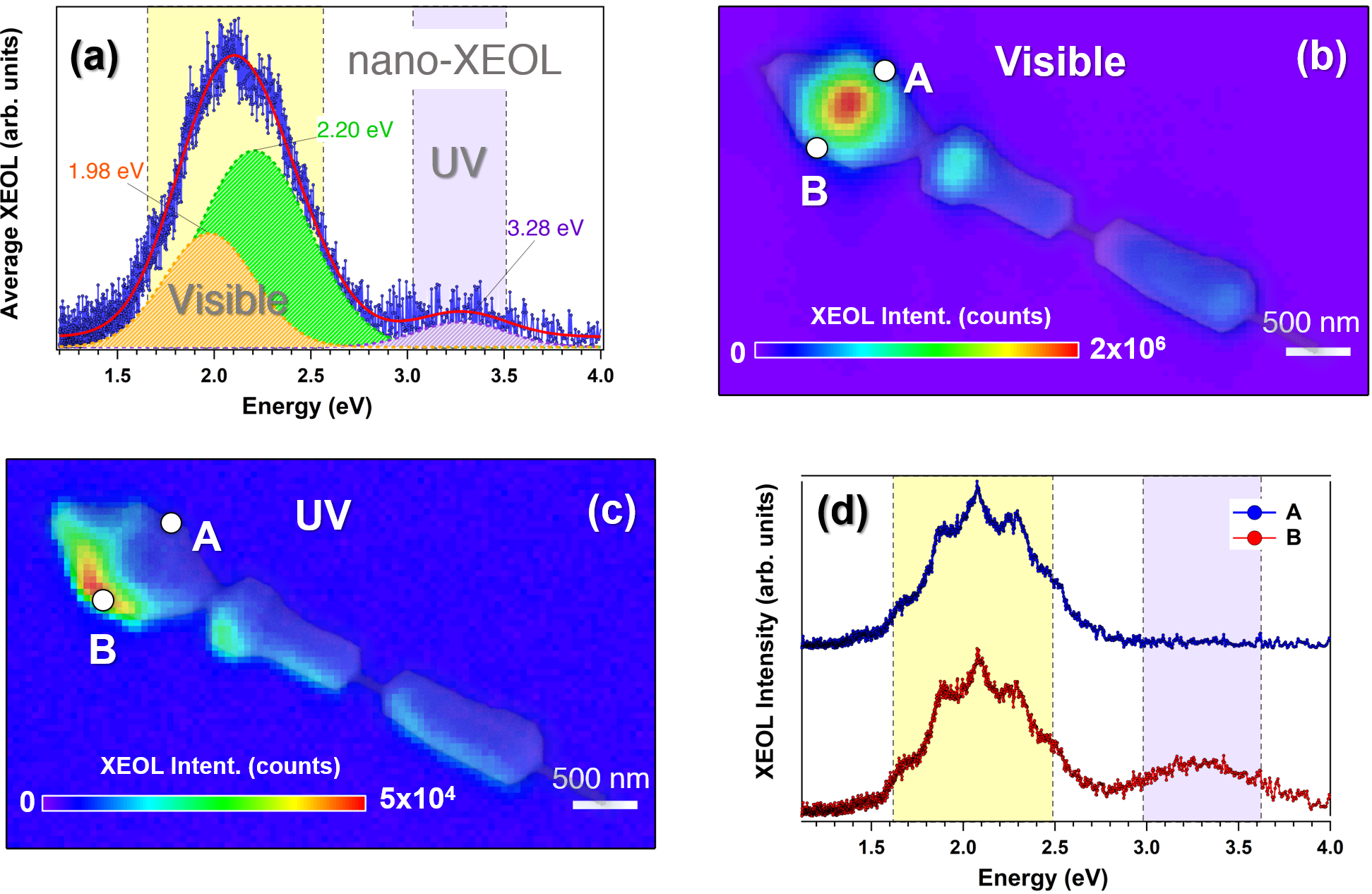}
        \caption{(a) Average XEOL spectrum recorded over the SK- structure shown in Figure 1a. (b) and (c) XEOL integrated intensity maps over the energy ranges defined by the colour bands in (a). (d) Normalized XEOL spectra taken from two different points (A and B).} 
        \label{fig2}
\end{figure}

We have also exploited the optical contrast mechanism to record spectrally and spatially resolved XEOL signals simultaneously to XRF data for each point of the sample (see Figure~\ref{fig2}). After the primary X-ray absorption process, most of the photogenerated electrons and holes recombine radiatively via secondary XRF and XEOL pathways. Figure~\ref{fig2}a presents the average XEOL spectrum, where a main emission dominates in the visible range, and consists of two components located at 1.98 and 2.20 eV (indicated in orange and green colours, respectively). In addition, there is a weak band in the UV range located at 3.28 eV. The main observed band in the average XEOL spectrum corresponds to the emission related to the transition of the trapped electrons from oxygen vacancies to intrinsic surface states from the Sn$_{1-x}$Ge$_x$O$_2$ crystallites. The luminescence of SnO$_2$ is quite complex, exhibiting a broad emission band in the visible range, consisting of three components (orange, green and blue at 1.98, 2.25, and 2.58 eV, respectively) associated to oxygen vacancies\cite{maestre2004, miller2017}. These components were also reported from SnO$_2$ nanoribbons by XEOL measurements excited with both soft and hard X-rays\cite{zhou2006, zhou2008}. Regarding the weak UV band, this emission could come from the Zn$_2$GeO$_4$ nanowire, which exhibits a wide emission associated to oxygen vacancies in the UV range\cite{dolado2020}. However, it has been observed that SnO$_2$ nanorods and nanoparticles exhibited an UV emission related to surface states\cite{kar2011}, so it could also come from Sn$_{1-x}$Ge$_x$O$_2$ particles.

Figures~\ref{fig2}b and ~\ref{fig2}c display the spatial distribution of the respective XEOL bands. The visible emission associated to oxygen vacancies is radially distributed within the Sn$_{1-x}$Ge$_x$O$_2$ crystallites, becoming particularly stronger from the well-faceted areas, where XRF signal of Sn is higher. Previously we observed that the crystallites of the SK-structures exhibit \{110\} facets\cite{dolado2019}, in which the formation of oxygen vacancies would be easy since they are the planes of lower surface energy in SnO$_2$\cite{oviedo2000}. Therefore, this result suggests that the higher amount of Ge in some areas of the Sn$_{1-x}$Ge$_x$O$_2$ particles would affect the amount of oxygen vacancies, with the consequent vanishing of the visible emission. On the other hand, the less intense UV emission of unclear origin is only present from one side of the Sn$_{1-x}$Ge$_x$O$_2$ crystallites. The fact that the UV emission recorded in this work is only observed from one side of the Sn$_{1-x}$Ge$_x$O$_2$ crystallite could suggest an orientation effect linked to the SnO$_2$ crystal structure. Previous work have reported faceted dependent cathodoluminescence in SnO$_2$ microtubes, which would imply different distribution of impurities and/or defects.\cite{maestre2004, yuan2006} Alternatively, the UV emission could come from the preferential incorporation of defects arising from the out-diffusion of Ge or Zn from the Zn$_2$GeO$_4$ nanowire, which would result in new electronic states within the bandgap leading to the UV emission.

In order to spatially resolve at the nanoscale these luminescence related features with more details, Figure~\ref{fig2}d shows the XEOL spectra acquired in two different points on the Sn$_{1-x}$Ge$_x$O$_2$   crystallites (A and B). In both cases, the visible band related to SnO$_2$ is the most pronounced contribution with some intensity modulation, resulting from the behaviour of the Sn$_{1-x}$Ge$_x$O$_2$ crystallites as optical cavities\cite{dolado2019}. On the other hand, the UV band is detected only from B, highlighting the inhomogeneous distribution of the radiative centres associated likely with the different distribution of impurities. As it is well known, the luminescence bands in semiconductors are related to the density of surface or defect electronic states\cite{kar2011, reshchikov2005}.

\begin{figure}
    \centering
        \includegraphics[width=1\textwidth]{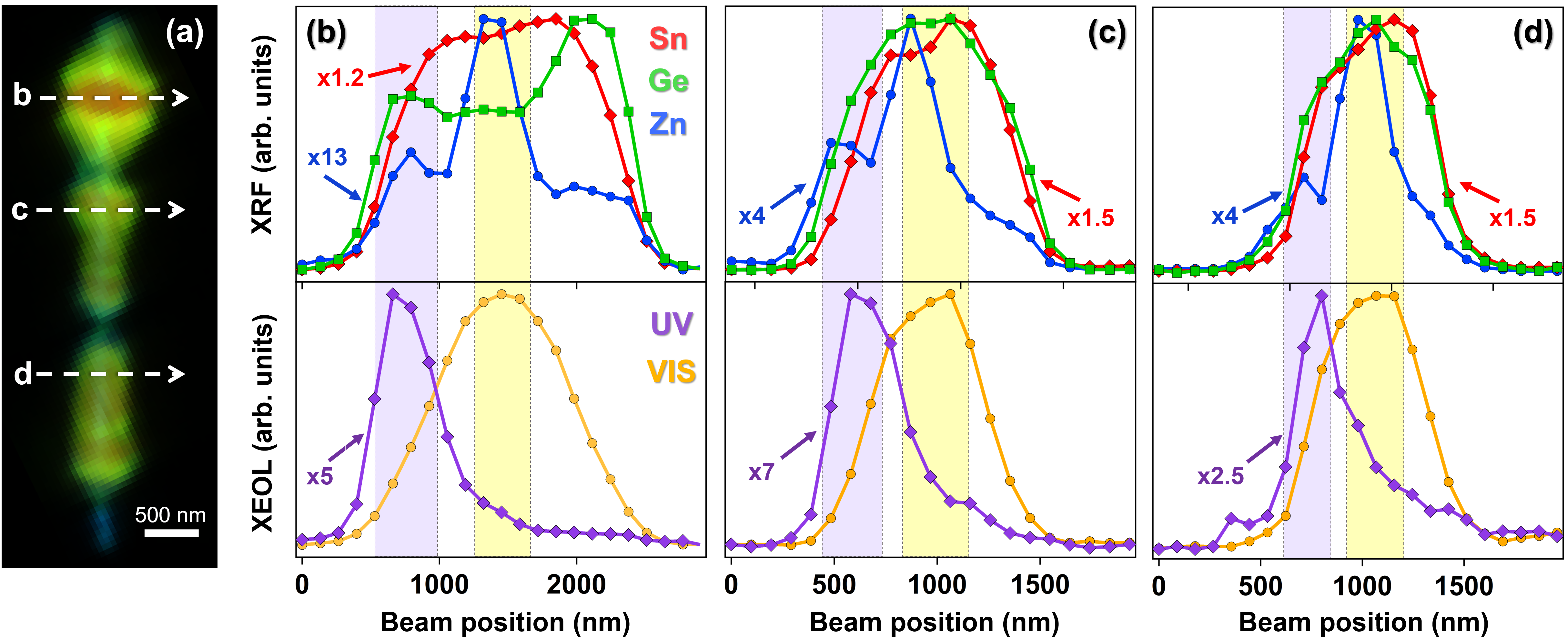}
        \caption{(a) SK-structure on which the line profiles were collected. (b-d) Normalized XRF line profiles (top) for Sn (red symbols), Ge (green symbols) and Zn (blue symbols), and XEOL line profiles (bottom) for visible (orange symbols) and UV (purple symbols) emissions. All profiles were collected along the white dashed lines drawn on the SK-structure shown on the left-hand side, perpendicular to the central nanowire.} 
        \label{fig3}
\end{figure}

In order to gain further information, given the heterogeneous nature of both composition and optical emissions, Figure~\ref{fig3} represents the normalized radial line profiles taken along the SK-structure (see Figure~\ref{fig3}a). Figures~\ref{fig3} b, c and d show the XRF and XEOL profiles data obtained from each of the three Sn$_{1-x}$Ge$_x$O$_2$ crystallites along the perpendicular direction to the central nanowire. For all cases, the Sn signal (in red colour) follows a similar trend according to the crystal volume with a maximum intensity at the centre of the Sn$_{1-x}$Ge$_x$O$_2$ crystallites. The Ge signal (in green colour), on the other hand, presents a different profile for the largest crystal compared to the other two ones. The largest crystallite appears to be fully formed, producing a higher intensity from the edges of the structure. The smaller crystallites are conversely characterized by a more uniform Ge incorporation, with a Gaussian like profile according to their morphological shape. Finally, the Zn profiles (in blue colour) display a multipeak shape, indicating a pretty different and particular behaviour within the crystallite volume. The results not only reveal the incorporation of Zn within the Sn$_{1-x}$Ge$_x$O$_2$ crystallites, but also its preferential spatial location towards the centre and edges of the crystallites.

Despite probing different sample depths (XEOL is much more surface sensitive than XRF), the radial line profiles of the XEOL signals provide complementary information about the three Sn$_{1-x}$Ge$_x$O$_2$ crystallites. Compared to the XRF signal, the spatial distribution of XEOL is not only governed mainly by the beam spot size and penetration of the X-ray nanobeam within the sample (energy dependent, leading to a finite generation volume), but also greatly by the diffusion of the photogenerated carriers. Many factors can impact the resulting profile related to the sample itself in terms of sample morphology, composition, internal absorption and/or reflection mechanisms, as well as secondary photon emissions. The visible band profile (in orange) follows the crystal volume with a Gaussian like shape well correlated with Sn [with a smaller full width at half maximum (FWHM) due to diffusion effects]. The profile of the UV emission (in violet) presents a quite narrower line width peaking at the left edge of the Sn$_{1-x}$Ge$_x$O$_2$ crystallite, matching one of the maxima in the Zn profile. The direct matching and interrelation between the UV band and Zn localization is also supported by axial line profiles recorded along the SK-structure (see Figure SI-3 in Supporting Information). It is important to emphasize that the UV emission is not observed in the bare Zn$_2$GeO$_4$ nanowire, ruling completely out their connection.

In short, the correlative study of the XRF and XEOL line profiles indicates that for the tin oxide crystallite fully formed (Figure~\ref{fig3}b), likely there is an additional colocalization of Ge and Zn elements. On the other hand, for all the cases (Figure~\ref{fig3} b, c, d) the UV emission is in close relationship with the Zn distribution mainly concentrated at one of the edges of the Sn$_{1-x}$Ge$_x$O$_2$ crystallites. This finding suggests that the UV emission could come from recombination centres linked to complexes associated to the presence of Zn.

\begin{figure}
    \centering
        \includegraphics[width=1\textwidth]{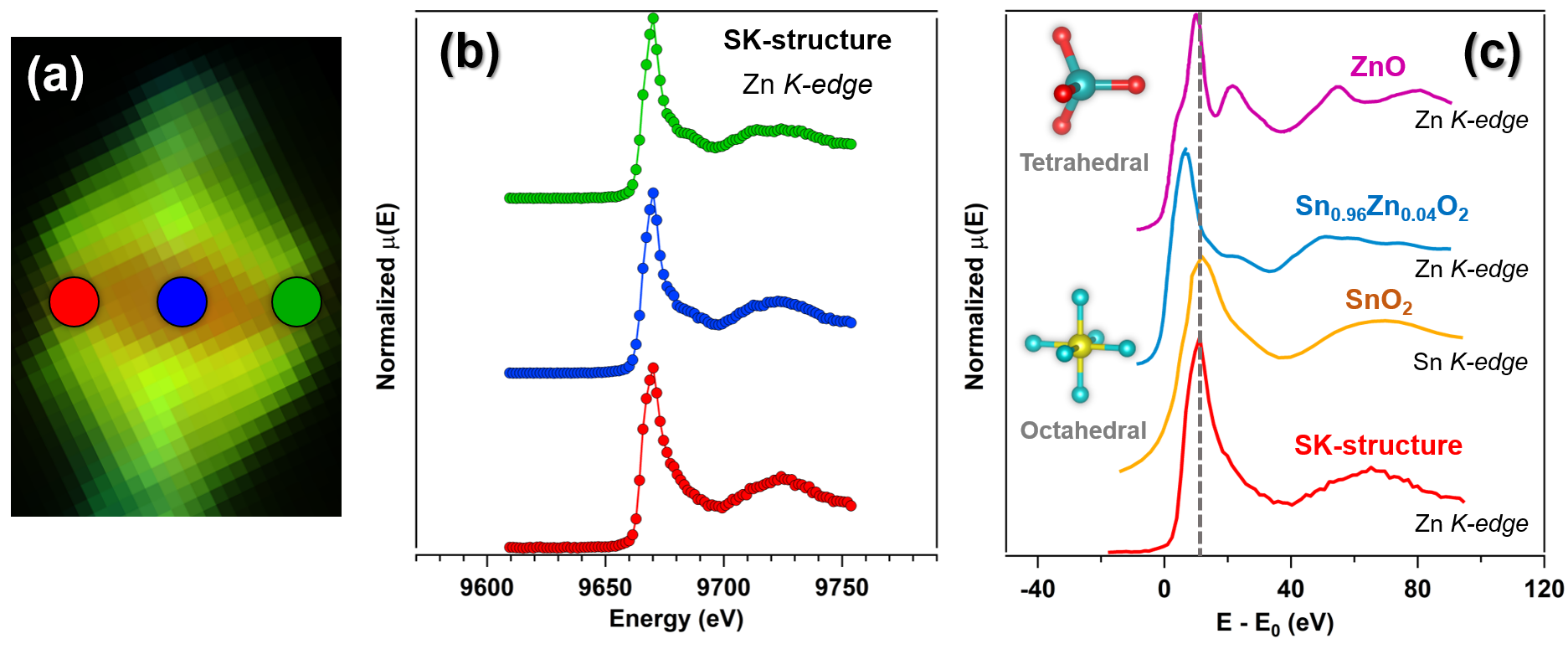}
        \caption{(a) Crystallite of the SK-structure in which the XANES spectra shown in (b) were acquired. Each spectrum was collected on the point of the same colour marked on the SK-structure. (c) XANES data recorded around the Zn or Sn K-edges of reference materials with tetrahedral (ZnO\cite{haug2011}), distorted octahedral (Sn$_{0.96}$Zn$_{0.04}$O$_2$\cite{liu2010}) and octahedral (SnO$_2$\cite{grandjean2004}) coordination, in addition to the one acquired in the SK-structure (at the bottom of the graph). The spectra were shifted vertically for clarity.} 
        \label{fig4}
\end{figure}

Given the key role of Zn within the Sn$_{1-x}$Ge$_x$O$_2$ crystallites, in order to get a deeper insight into its local structure in the SK-structure, XANES data were taken around the Zn K-edge with nanometer spatial resolution. The measurements were performed with a beam size of 80 x 60 nm$^2$ using a Si(111) double crystal monochromator with an energy resolution about 1 eV. Figure~\ref{fig4}b shows the spatially resolved XANES data acquired in XRF detection mode over three different zones of the SK-structure. The XANES signal is analogous to the partial density of the empty states of the absorbing atoms; namely, to the energy band structure for crystalline material. In particular, XANES collected at the K edge reveals the $p$-partial density of states in the conduction band. Here, two spectral XANES regions can be distinguished: a prominent white line assigned mainly to the 1$s$ → 4$p$ dipolar transitions, and the post-edge resonances, which are rather due to multiple scattering effects. The three XANES spectra collected on the Sn$_{1-x}$Ge$_x$O$_2$ crystallite show an intense and narrow peak at 9670.2 eV, and the post-edge are similar in all three cases, indicating a similar short-range order around Zn atoms in the different areas.

An empirical and common practice is to assume that the XANES spectrum from an unknown local structural order can be understood as a linear superposition of the spectra of two or more known samples. Therefore, for comparison XANES spectra around the Zn or Sn K-edges of three model compounds are plotted in Figure~\ref{fig4}c, in addition to one of those acquired from the Sn$_{1-x}$Ge$_x$O$_2$ crystallites. In ZnO\cite{haug2011}, Zn atoms are in tetrahedral sites. In the Sn$_{0.96}$Zn$_{0.04}$O$_2$ alloy, on the other hand, Zn is introduced into the SnO$_2$ structure, inducing numerous lattice defects with no big change of the rutile crystal structure (octahedral environment)\cite{liu2010}. Finally, the XANES spectrum around Sn K-edge acquired on SnO$_2$ with rutile structure is shown\cite{grandjean2004}, where each tin atom is expected to be coordinated by six oxygen atoms. Our results strongly suggest a substitutional Zn incorporation into the Sn sites. A direct comparison of XANES data from the post-edge region around the Zn and Sn K-edges for the Sn$_{1-x}$Ge$_x$O$_2$ and SnO$_2$, respectively, indicates the predominant octahedral coordination of Zn atoms in the Sn$_{1-x}$Ge$_x$O$_2$ structure. In addition, theoretical XANES simulations of ZnO in a rutile-like environment show similar spectral shapes, corroborating our results\cite{zhu2014}. Thus, the fact that the XANES spectrum of the Sn$_{1-x}$Ge$_x$O$_2$ crystallites is significantly different from the XANES data of the Sn$_{0.96}$Zn$_{0.04}$O$_2$ alloy suggests that Zn is incorporated as a dopant into the rutile structure, without causing major lattice distortions. In addition, the white line of the SK-structure is shifted to higher energies with respect to those of ZnO and the Sn$_{0.96}$Zn$_{0.04}$O$_2$ alloy, indicating an increase in Zn coordination.

In short, XANES results point to Zn atoms in octahedral coordination within the Sn$_{1-x}$Ge$_x$O$_2$ crystallites, indicating that it is being incorporated into the crystal lattice possibly as a dopant replacing Sn in the rutile structure. This substitutional Zn defect in the SnO$_2$ lattice may lead to an electronic state within the gap responsible for the UV emission above observed by XEOL.

In summary, we showed the use of a hard X-ray nanoprobe with optical, elemental and local structure specificity for the study of Zn$_2$GeO$_4$/SnO$_2$ heterostructures grown by an effective thermal evaporation method. Both the compositional and optical distributions were probed with nanometre resolution. Elemental maps showed that Ge and Zn atoms diffused from the central Zn$_2$GeO$_4$ nanowire to the Sn$_{1-x}$Ge$_x$O$_2$ crystallite with an inhomogeneous character. On the other hand, XEOL maps revealed a homogeneous repartition of the dominant SnO$_2$ related visible emission whereas the UV emission is only present on one side of the crystallite. A correlative study between XRF and XEOL line profiles allowed us to associate the presence of the UV band with a higher Zn concentration in certain areas of the Sn$_{1-x}$Ge$_x$O$_2$ crystallite. XANES data collected in different areas of the crystallite around the Zn K-edge confirmed possible Zn octahedral coordination in the crystallites of the SK-structure. This finding could likely indicate the substitutional incorporation of Zn into the Sn sites in the rutile structure, leading to an electronic state within the bandgap responsible for the UV emission. Thus, this work shows how to complete the picture through correlative X-ray nano-analysis and how to develop wide bandgap nanowire-based architectures with bespoke properties at the nanoscale.

\section{Associated content}

Both distribution maps and normalized line profiles collected along the SK-structure from XRF and XEOL signals were processed with PyMca software.

\begin{acknowledgement}

This work has been supported by MICINN project RTI2018-097195-B-I00 and by European Union’s Horizon 2020 research and innovation programme under the Marie Skłodowska-Curie grant agreement n° 956548, project Quantimony. We thank the ESRF for the beam time allocated. 

\end{acknowledgement}

\begin{suppinfo}

Vertical and horizontal X-ray beam profiles; Sn, Ge and Zn XRF maps of the SK-structure; and normalized XRF and XEOL line profiles collected along the SK-structure parallel to the central nanowire. This material is available free of charge.

\end{suppinfo}

\bibliography{references}

\end{document}